# Dots and Boxes Algorithm for Peierls Substitution: Application to Multidomain Topological Insulators

Ricardo Y. Díaz-Bonifaz and Carlos Ramírez*

Departamento de Física, Facultad de Ciencias, Universidad Nacional Autónoma de México, Apartado Postal 70542, Ciudad de México 04510, México

*Corresponding author e-mail address: carlos@ciencias.unam.mx

**Abstract**

Magnetic fields can be introduced into discrete models of quantum systems by the Peierls substitution. For tight-binding Hamiltonians, the substitution results in a set of (Peierls) phases that are usually calculated from the magnetic vector potential. As the potential is not unique, a convenient gauge can be chosen to fit the geometry and simplify calculations. However, if the magnetic field is non-uniform, finding a convenient gauge is challenging. In this work we propose to bypass the vector potential determination by calculating the Peierls phases exclusively from the gauge-invariant magnetic flux. The phases can be assigned following a graphic algorithm reminiscent of the paper and pencil game "*dots and boxes*". We showcase the method implementation by calculating the interference phenomenon in a modified Aharonov-Bohm ring and propose a phase assignation alternative to the Landau gauge to reproduce the Half Integer Quantum Hall Effect in graphene. A non-uniform magnetic field case is addressed by considering a multi-domain Chern insulator to study the effects of domain walls in resistance and current quantization. It is found that adding decoherence and a finite temperature into the model results in quantized resistances that are in good agreement with experiments made with multi-domain intrinsic topological insulators.

1.  **Introduction**

Magnetic fields in quantum systems are a source of interesting phenomena with great technological applicability. For instance, the discovery of the quantum Hall [1] and Aharonov-Bohm [2] effects triggered decades of research in condensed matter physics where magnetic fields play a key role [3–11]. The presence of magnetic fields in solids is also an important feature in the study of topological matter, in which different phases are characterized by topological invariants [12]. These systems are of utmost interest due to the appearance of robust edge states with properties such as absence of backscattering [13] or charge fractionalization [14] that depend only on the value of the topological invariant that remains unaltered by adiabatic transformations. Applying a magnetic field to a solid is a simple way to break the time reversal symmetry and is related to the appearance of non-trivial topological phases. In particular, the quantum Hall effect is related to the Chern number or TKNN invariant, which is an invariant that can be modified by an applied magnetic field [15].

An interesting case involving topological insulators (TIs) occurs when a system is constituted by multiple topological phases, which is the case of the embedded TIs [16] and the multidomain TIs [6,17–19]. The frontiers between different topological domains in multidomain TIs are known as domain walls (DWs), which have been proposed for magnetic memories [20], nanoelectronics [21], and quantum information transfer [22]. Thus, in multidomain Chern insulator the local Chern number, that indicates the number of edge states due to the bulk-boundary correspondence, is not uniform within the sample. As the Chern number is related to the magnetic field, a multidomain Chern insulator might be formed by applying a non-uniform magnetic field. Multidomain Chern insulators have been successfully synthesized utilizing intrinsic topological insulators. This has allowed to describe the transport properties when domain walls are present [6], as well as to develop techniques to synthesize multidomain systems with arbitrary Chern number difference between adjacent domains [18,19].

The analysis of TIs is often made in k-space [23], which results inconvenient for multidomain systems as the DWs break the periodicity and require real space models that do not restrict the system to be periodic. A way in which transport properties have been theoretically estimated in multidomain Chern insulators is by using the Landauer-Büttiker formula [24]. This formula is given in terms of transmission coefficients, which can be calculated from real space models to include the effects of disorder, decoherence or edge state interactions.

Let us focus on quantum systems described by tight-binding Hamiltonians, where we have a basis of localized functions represented by the set of kets $\{|n\rangle\}$. In this basis, the Hamiltonian operator is given as

$$\hat{H} = \sum_n \left( \varepsilon_n |n\rangle\langle n| + \sum_{n \neq m} t_{n,m} |m\rangle\langle n| \right), \tag{1}$$

where the terms $\varepsilon_n = \langle n|\hat{H}|n\rangle$ and $t_{n,m} = \langle m|\hat{H}|n\rangle$ are on-site energies and hopping integrals respectively. The sums develop over all the sites in the lattice. The effect of a static magnetic field can then be introduced by modifying the momentum of the electrons as $\mathbf{p} \rightarrow \mathbf{p} + \frac{e}{c}\mathbf{A}(\mathbf{r})$, where $e$ is the electron charge and $c$ is the speed of light [25,26]. The vector potential $\mathbf{A}(\mathbf{r})$ relates to the magnetic field as $\mathbf{B} = \nabla \times \mathbf{A}$. As a result of the substitution, the hopping integrals take the form $t_{n,m} = t_0 \exp(-i\theta_{n,m})$, where $t_0$ is the hopping integral when there is no magnetic field and $\theta_{n,m}$ is the Peierls phase (PP) that the particles acquire when they travel from the $n$-th into the $m$-th lattice site. This phase can be calculated in terms of the magnetic vector potential as [25]

$$\theta_{n,m} = \frac{2\pi}{\Phi_0} \int_{\mathbf{R}_n}^{\mathbf{R}_m} \mathbf{A}(\mathbf{r}) \cdot d\mathbf{r} \tag{2}$$

where $\Phi_0 = \frac{2\pi\hbar}{e}$ is the magnetic flux quantum. The potential $\mathbf{A}(\mathbf{r})$ is not gauge invariant, since

$$\mathbf{A}(\mathbf{r}) \rightarrow \mathbf{A}(\mathbf{r}) + \nabla \chi(\mathbf{r}) \tag{3}$$

give us the same magnetic field. In consequence,

$$\theta_{n,m} \rightarrow \theta_{n,m} + \frac{2\pi}{\Phi_0}(\chi_m - \chi_n), \tag{4}$$

*i.e.*, PPs are non-gauge invariant. On the other hand, the magnetic flux $\Phi_S$ through a surface $S$ encircled by the closed loop $\partial S$ is

$$\Phi_S = \int_S \mathbf{B} \cdot d\mathbf{S} = \oint_{\partial S} \mathbf{A} \cdot d\mathbf{r} \tag{5}$$

where the second equality comes from Stokes' theorem. Gauge invariance of $\mathbf{B}$ implies that $\Phi_S$ is gauge invariant as will be the vector potential integral over any closed path.

Finding a convenient gauge might be a challenging task even for uniform magnetic fields. The urge to describe complex lattices with promising applications such as twisted multilayer graphene has led to recent developments in system-specific simplifications of the PP calculation [27,28]. Moreover, gauge selection is not trivial whenever there are infinite leads, as well as for 2D infinite systems. For instance, let us consider the nanoribbon shown in Fig. 1(a) with a uniform magnetic field perpendicular to its surface. The nanoribbon is infinite and periodic in the x direction with lattice parameter $a$ and a constant magnetic flux $\Phi$ per unit cell. The Landau gauge $\mathbf{A} = \left(-\frac{x\Phi}{a^2}, 0, 0\right)$ produces PPs that do not differ between different unit cells and permits the use of Bloch theorem because periodicity is maintained. Let us now consider the system presented in Fig. 1(b), in which there are differently oriented leads. A vector potential must be found so that the PPs do not change between unit cells in the leads. This can be achieved by identifying piece-wisely a convenient gauge for each part of the system. Let us call $\mathbf{A}_i(\mathbf{r})$ the desired potential for the $i-$th lead. For the system to be described adequately it is necessary to define $\chi(\mathbf{r})$ in eq. (3) so that $\mathbf{A}(\mathbf{r})$ becomes $\mathbf{A}_i(\mathbf{r})$ in the $i-$th lead. As $\chi(\mathbf{r})$ must be differentiable, some authors have proposed smooth transitions between the different regions [29–31], resulting in complicated expressions that are not easily generalized. This was further simplified by A. Cresti arguing the fact that $\chi(\mathbf{r})$ may vary significantly in distances shorter than the lattice parameter, which means that the smooth evolution of $\mathbf{A}(\mathbf{r})$ is not necessary for discrete systems [32]. This resulted in a formula to calculate convenient PPs that are suitable for differently oriented leads under uniform magnetic fields.

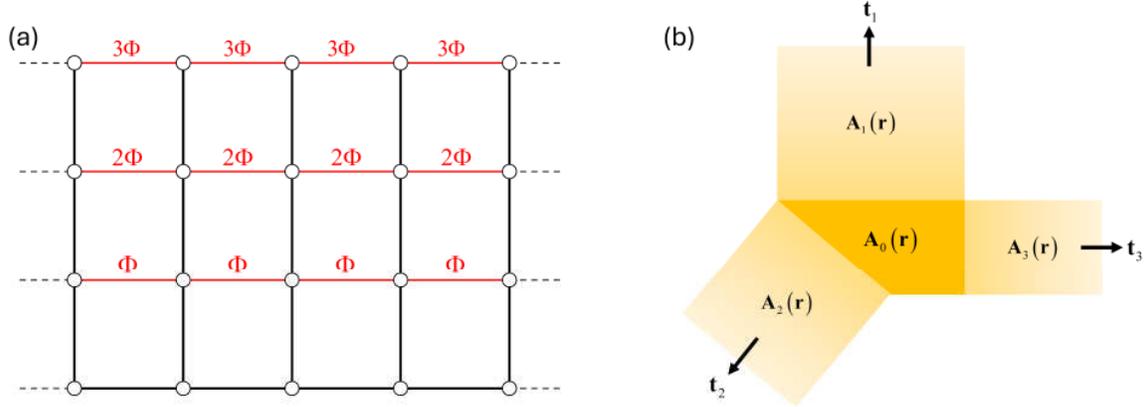

**Figure 1:** (a) Peierls phases calculated from a Landau gauge calculated for an infinite periodic nanoribbon. Black and red lines represent null and non-zero phases respectively. (b) Schematic representation of a system with leads in different directions.

Despite the recent developments in simplifying the PP calculation, it is still necessary to develop techniques that are suitable for non-uniform magnetic fields, where an explicit expression for $\mathbf{A}(\mathbf{r})$ might be unknown or inconvenient. In the present paper we address this challenge by exploiting the lack of gauge invariance of the vector potential to develop a graphic algorithm reminiscent to the "*dots and boxes*" pencil and paper game that is suitable for both uniform and non-uniform magnetic fields. The algorithm is described in detail in section 2. In section 3 we show how the algorithm can be implemented to study a modified Aharonov-Bohm ring and to simplify the calculations of the resistances for the quantum Hall effect in graphene. A non-uniform magnetic field is considered in section 4 to form a multi-domain Chern insulator, calculating the resistance between different terminals and show the role of decoherence and finite temperature to achieve an agreement with experimental results.

## 2. Dots and Boxes Algorithm for Peierls Substitution

In this section we present an algorithm to assign Peierls phases in discrete systems directly from the gauge invariant magnetic flux without the explicit calculation of the vector potential.

To that end, let us consider a set of $N$ fixed lattice points $\{\mathbf{r}_n\}$ and a closed simple path $C$ formed by straight lines that connect sequentially $\mathbf{r}_n$ and $\mathbf{r}_{n+1}$ with $\mathbf{r}_{N+1} = \mathbf{r}_1$, as shown in Fig. 2. In a tight-binding description we can think that every line that connects two different sites represents a hopping parameter for which the PP needs to be calculated. The total PP $\theta_C$ acquired throughout the path $C$ is

$$\theta_C = \frac{2\pi}{\Phi_0} \oint_C \mathbf{A}(\mathbf{r}) \cdot d\mathbf{r} = 2\pi \frac{\Phi_C}{\Phi_0}, \tag{6}$$

where Eq. (5) was used and $\Phi_C$ is the total magnetic flux through the surface delimited by $C$. By defining $\theta_{n,n+1}$ as the PP acquired by traveling through $C$ between the points $\mathbf{r}_n$ and $\mathbf{r}_{n+1}$ we get that

$$\theta_C = \sum_{n=1}^{N} \theta_{n,n+1}. \tag{7}$$

In a closed path involving $N$ points there will be an equal number of PPs to be determined. Notice from Eq. (4) that the phases $\theta_{n,n+1}$ are not gauge invariant themselves and Eq. (6) represents the only physically relevant constraint for the values they can take.

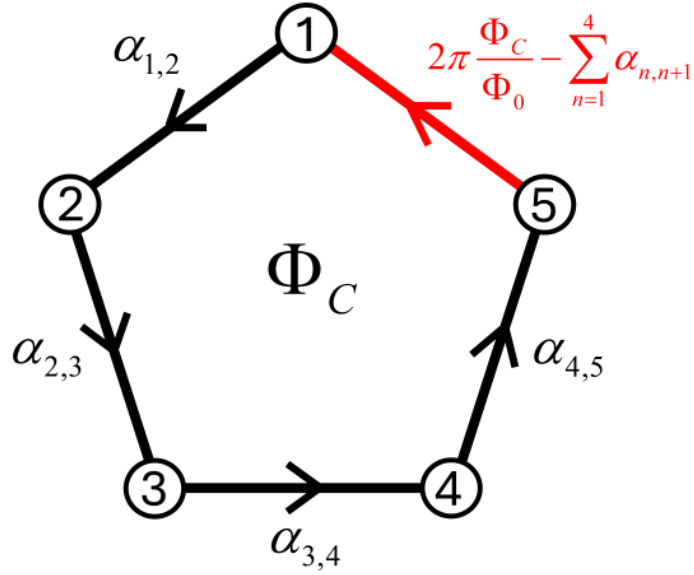

**Figure 2:** Simple closed path that goes through $N=5$ lattice points. The terms $\alpha_{n,n+1}$ are arbitrary phases and the red term denotes the required value for Eq. (10) to be satisfied.

We will now show that we can set $N-1$ of the PPs to have arbitrary values unrelated with the magnetic field without losing consistency with Eq. (2). Let us assume that we have a static magnetic field (no necessarily uniform) that can be obtained from a vector potential $\mathbf{A}^0(\mathbf{r})$ which can be substituted in Eq. (2) to get a set of PPs $\theta^0_{n,n+1}$. We can then transform the potential by adding the gradient of a function $\chi(\mathbf{r})$ as in Eq. (3) that evaluated in the lattice points takes the values $\chi(\mathbf{r}_n) = \chi_n$. Notice that in a discrete system the distance between any pair of lattice points is finite and we can always think of a continuous and differentiable function $\chi(\mathbf{r})$ that take any set of values $\{\chi_n\}$ that we want. In other words, the values of $\chi_n$ are arbitrary. By introducing $\chi(\mathbf{r})$ into Eq. (4), the PP calculated between any pair of consecutive lattice points will be given as

$$\theta_{n,n+1} = \theta^0_{n,n+1} + \frac{2\pi}{\Phi_0}(\chi_{n+1} - \chi_n). \tag{8}$$

If we want to set the first $N-1$ PPs to have arbitrarily chosen values $\theta_{n,n+1} = \alpha_{n,n+1}$ for $n \leq N-1$, we can substitute into equation (8) and obtain

$$\chi_{n+1} = \frac{\Phi_0}{2\pi}\left(\alpha_{n,n+1} - \theta^0_{n,n+1}\right) + \chi_n \tag{9}$$

as the recurrence rule for $\chi_{n+1}$. Notice from Eq. (9) that we are still free to keep $\chi_1$ as a free parameter. The last PP must be set to be

$$\theta_{N,1} = 2\pi \frac{\Phi_C}{\Phi_0} - \sum_{n=1}^{N-1} \alpha_{n,n+1}, \tag{10}$$

which guarantees that our gauge invariant condition given in Eqs. (6) and (7) can always be satisfied.

We can conclude that when calculating PPs between lattice sites in a way that forms a simple closed loop we are free to assign any arbitrary value to all the PPs except for the last one, which must be calculated from Eq. (10). An important aspect of the previous discussion is that it is also valid for non-uniform magnetic fields, since the only relevant restriction is given by Eq. (6) and it only depends on $\Phi_C$. This means that it is not necessary to calculate any vector potential to assign PPs. In the following, we synthesize this argument in the form of a graphic algorithm that we call "*dots and boxes*" as it is reminiscent of the paper and pencil game that consists of drawing lines between points in a lattice to form closed boxes until the lattice is complete [33].

### 2.1 Dots and Boxes Algorithm for Peierls Substitution

Let us consider an arbitrary system described by a tight-binding Hamiltonian (Eq. (1)) given in terms of a basis of localized Wannier functions represented by the kets $\{|n\rangle\}$, whose centers form a lattice. We will refer to every pair of sites with non-zero hopping parameter as connected sites and represent a "bond" as a straight line that links two connected sites. To introduce a magnetic field into the Hamiltonian we must determine a PP for every bond. The dots and boxes algorithm for PP assignation can be summarized in the following steps:

1. Select any pair of connected sites and draw the bond between them as a straight line.
2. If the line does not create a new box (closed region) in the lattice, you are free to assign any value you find convenient to the PP associated to that bond. However, if the line creates a new box, the phase assigned to the bond must be such that Eq. (10) is satisfied.
3. Repeat steps 1 and 2 for different pairs of connected sites until the PPs are determined for all the desired connections.

To clarify the procedure, let us apply the dots and boxes algorithm to the lattice shown in Fig. 3(a), where the dashed lines indicate the desired bonds. We begin by focusing on sites $1$ and $2$. As this is the first bond we draw, it does not form a box and we can choose any arbitrary value for the PP. For simplicity, we will assign a zero phase to every bond that does not form a box, so $\theta_{2,1} = 0$. Thus, if we continue the assignation with $\theta_{3,2}$, $\theta_{6,3}$, $\theta_{5,6}$ and $\theta_{1,4}$ we can make them all zero as no box is formed by drawing those bonds. However, by adding the bond between $4$ and $5$ we form the box shown in Fig. 3(b). Considering a counter-clockwise convention for phase acquisition, Eq. (10) results in $\theta_{1,4} = \theta_A = \frac{2\pi \Phi_A}{\Phi_0}$, where $\Phi_A$ is the flux through the box $A$ shown in Fig. 3(b).

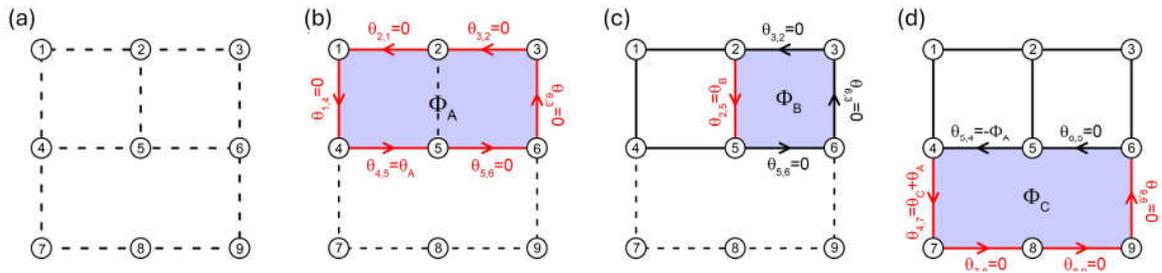

**Figure 3:** Schematic representation of a lattice in which the Peierls phases are assigned through the dots and boxes algorithm. (a) shows the desired connections as dashed lines while (b)-(d) show three different boxes that allow to complete the phase assignment.

If we then draw the bond between 2 and 5 we form the box $B$ shown in Fig. 3(c) and Eq. (10) implies that $\theta_{2,5} = \theta_B = \frac{2\pi\Phi_B}{\Phi_0}$, where $\Phi_B$ is the magnetic flux through the box $B$. We can then assign $\theta_{9,6} = \theta_{8,9} = \theta_{7,8} = 0$ until we reach the bond between sites 4 and 7 which forms the box $C$ that has a flux $\Phi_C$ as shown in Fig. 3(d). Notice that the only known non-zero phase associated to the bonds that form box $C$ is $\theta_{5,4} = -\theta_{4,5} = -\frac{2\pi\Phi_A}{\Phi_0}$, so by substituting in Eq. (10) we have that $\theta_{4,7} = \theta_A + \theta_C = \frac{2\pi}{\Phi_0}(\Phi_C + \Phi_A)$ and all the desired PPs are finally known. It is worth mentioning that the set of phases we just calculated is not unique and there are infinite different sets that can be reached by changing the order in which the connections are drawn or by choosing different phases for the connections that do not form a box.

## 3. Examples of algorithm implementation

In this section we calculate simple and convenient PPs to obtain numerically the interference in an Aharonov-Bohm (AB) ring and the half-integer quantum Hall effect (HIQHE) in graphene. These effects have been studied thoroughly in the past [2,5,10,11,34]. However, to expose the advantages provided by the dots and boxes algorithm, we propose a variation of the Aharonov-Bohm ring so that the total flux is divided in two regions. For the graphene Hall bar, a PP assignation is proposed that does not correspond to a Landau gauge. The aim of the present section is to serve as validation of the method while presenting new results in well-known phenomena.

### 3.1 Aharonov-Bohm interference

Let us consider an AB ring formed by $N$ sites as shown in Fig. 4(a). The ring is provided with two infinite chains (leads) connected to the sites $N/2$ and $N$. It is a well-known result that the transmittance between the leads will depend on the magnetic flux $\Phi_R$ through the ring in periodic fashion with period $\Phi_0$ [2]. To recover this result from a discrete model, we can describe the AB ring and the leads with a tight-binding Hamiltonian (1) with constant hopping parameter $t_0$. It is customary to introduce the magnetic field into the Hamiltonian by assigning a constant PP

$$\theta_{n,n+1} = \frac{2\pi \Phi_R}{\Phi_0 N} \quad (11)$$

to every bond within the ring. This choice, which is translationally invariant within the ring, is consistent with the dots and boxes algorithm. We refer to Eq. (11) as the *uniform phase* approach. Another option is to calculate the PPs through Eq. (2) by taking the Landau gauge $\mathbf{A}(\mathbf{r}) = \mathbf{A} = \left(0, \frac{x\Phi_R}{A_R}, 0\right)$, for which the phases vanish in the leads. The term $A_R$ is the area delimited by the ring. This gauge leads to PPs of the form

$$\theta_{n,n+1} = \frac{\pi \Phi_R}{A_R \Phi_0} (y_{n+1} - y_n)(x_{n+1} + x_n), \quad (12)$$

where $(x_n, y_n)$ indicates the position of the $n$-th lattice site.

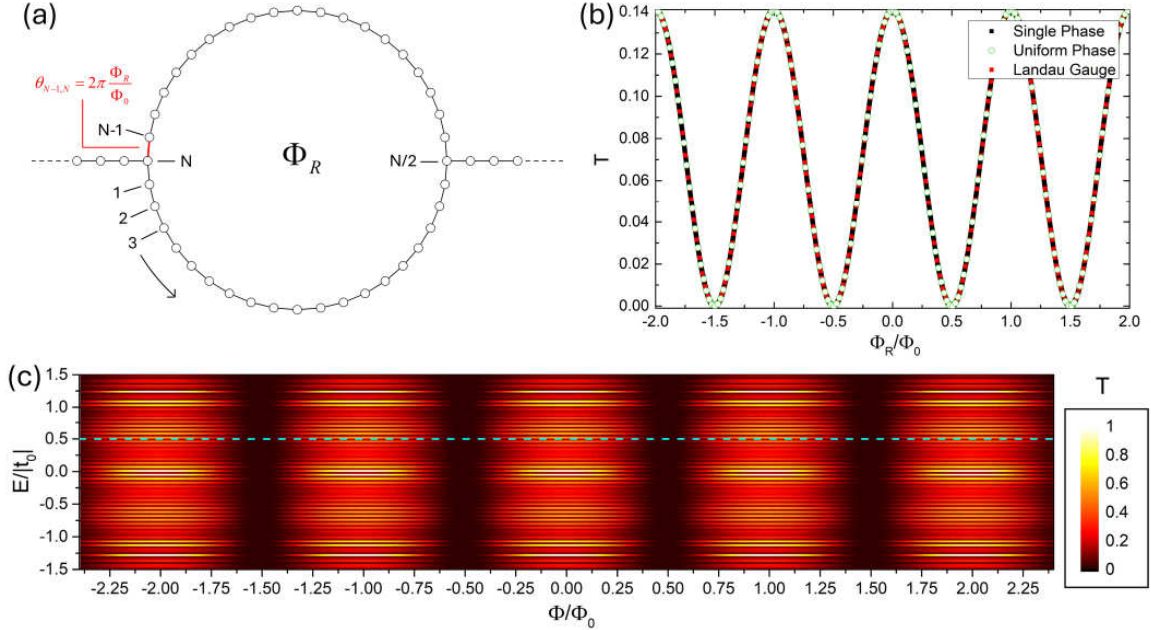

**Figure 4:** (a) Schematic representation of the single-phase Peierls phase assignation in an $N-$site long Aharonov-Bohm ring with two infinite leads. The red line denotes the only hopping with non-zero phase. (b) and (c) Left to right transmittance as a function of the magnetic flux $\Phi_R$. In (b) different forms of assigning the phases are compared for a fixed energy $E = 0.5|t_0|$. The results in c. are calculated from the single phase approach when the energy is varied.

Instead of the previous proposals, we can follow the dots and boxes algorithm to set all the phases within the ring to zero except for the last one $\theta_{N-1,N}$, which from Eq. (10) is set to be

$$\theta_{N-1,N} = 2\pi \frac{\Phi_R}{\Phi_0}, \tag{13}$$

as shown in Fig. 4(a). This later version, which we will refer to as the *single phase* approach, is similar to the one utilized in Ref. [5] which was not obtained through the dots and boxes algorithm but by considering a sharp vector potential $\mathbf{A}(\mathbf{r}) = \left(\frac{\Phi_R}{\Phi_0} 2\pi \theta(-y)\delta(x), 0, 0\right)$, which leads to a single bond with non-zero PP within the ring. It is worth mentioning that following the dots and boxes algorithm the PPs in the infinite chains that form the leads can take arbitrary values as they never form a closed loop, we hereby make those phases zero for convenience.

We calculate the S-matrix of this system following the recursive S-matrix method (RSMM) and the transmittance can be calculated following the procedure described in references [35] and [36]. In Fig. 4(c) we show the transmittance calculated from the single phase PP assignation as a function of energy and magnetic flux for an $N = 198$ sites long AB ring. Observe that the transmittance has a periodicity of $\Phi_0$ as $\Phi_R$ varies. Moreover, in Fig. 4(b) we compare the AB transmittance oscillations calculated from the single phase approach with the ones calculated with from Eq. (11) (uniform assignation) and with Eq. (12) (Landau gauge), resulting in the same pattern. This confirms that the three forms of PP assignation are completely equivalent.

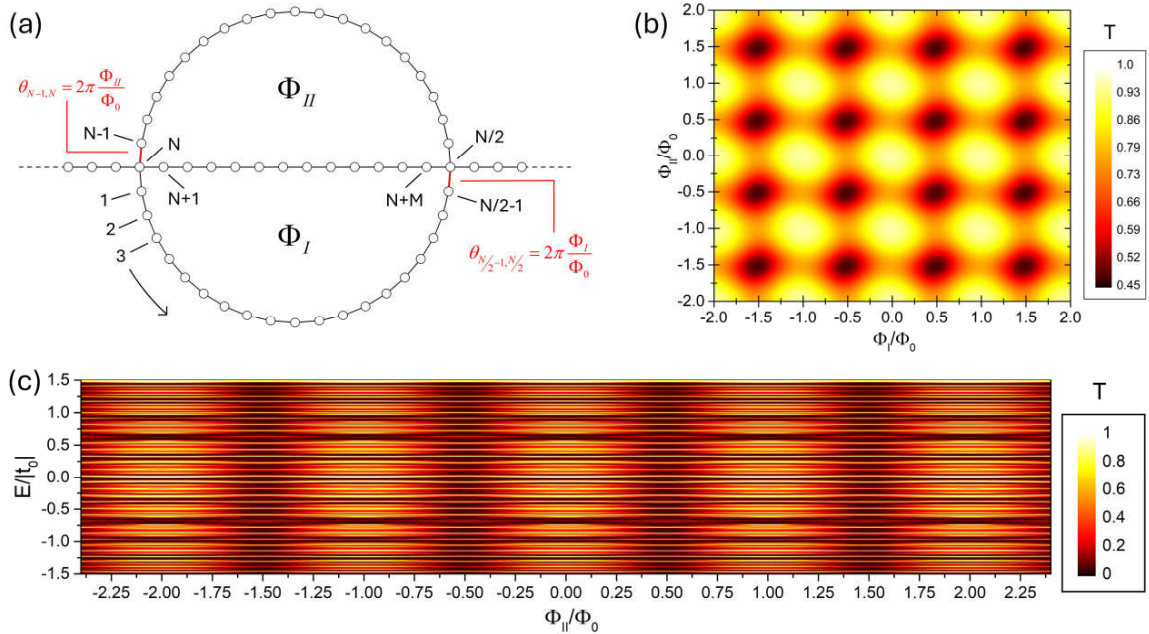

**Figure 5:** (a) Divided Aharonov-Bohm ring where the bonds with non-zero PP are shown in red. (b) Left to right transmittance as a function of the magnetic fluxes $\Phi_I$ and $\Phi_{II}$ for a fixed energy of $E = 1.5|t_0|$. (c) Left to right transmittance calculated for the case where there is no magnetic field in the lower portion of the ring, *i.e.* for $\Phi_I = 0$.

To exploit the algorithm even further we can calculate the transmittance through an AB ring when an $M-$site long chain divides the area within the ring in two regions with fluxes $\Phi_I$ and $\Phi_{II}$ respectively. The proposed divided AB ring is presented in Fig. 5(a). This simple modification to the original AB ring represents a challenge to calculate the PPs as we would have to find a vector potential that results in the correct magnetic flux in each region. However, we can avoid this inconvenience by following the dots and boxes algorithm. Notice that we now have a system with two new possible closed paths. Following the dots and boxes algorithm we can choose only two bonds, one from the lower portion of the ring and the other from the upper portion, and set their phases to be $2\pi\Phi_I/\Phi_0$ and $2\pi\Phi_{II}/\Phi_0$ respectively. The rest of the bonds, including the ones from the chain in the center, are set to have null PP. This approach is shown schematically in Fig. 5(a). Notice that we can choose any closed loop in the divided ring and the phase summation satisfies Eq. (10).

In Fig. 5(b) we show the calculated transmittance as a function of both $\Phi_I$ and $\Phi_{II}$ for a fixed energy of $E = 3|t_0|$ when an $M = 62$ site long chain is considered in the center. Observe that whenever $\Phi_I$ and $\Phi_{II}$ equal an integer multiple of $\Phi_0$ the transmittance has a maximum, which means that the AB interference is preserved despite the chain in the center. On the other hand, in Fig. 5(c) we show the transmittance as a function of $E$ and $\Phi_{II}$ by setting $\Phi_I = 0$. Notice how in this case the $\Phi_0$ periodicity is preserved from the original AB ring, but the energy dependance is modified from the one shown in Fig. 4(c). This can be understood as the lower portion of the ring preserves the phase with which the electron arrives from the lead and modifies the interference pattern.

### 3.2 Half-Integer Quantum Hall Effect in Graphene

In the following the dots and boxes algorithm is applied to calculate resistances in a graphene multiterminal setup to obtain the half-integer quantum Hall effect (HIQHE) [10,11,34]. Let us consider a rectangular shaped graphene sheet with six infinite leads attached, as shown in Fig. 6. Notice that the system is such that the left and right borders are in a zigzag configuration while the up and down borders are armchair. We define $\Phi_H$ as the magnetic flux through any hexagon in the lattice. Following the dots and boxes algorithm, a simple way to set the PPs in the Hall bar (excluding the leads) is by setting all the phases to zero except for the ones that unite the sites horizontally. The PPs associated to horizontal bonds can be set to evolve in integer multiples of $\theta_H = 2\pi \frac{\Phi_H}{\Phi_0}$ as we move upward. This is shown by the red lines in the Figs. 6(b) and 6(d). It is worth mentioning that this assignation does not correspond to any Landau gauge. This same technique can be applied to leads $L_1$ and $L_4$ (see Fig. 6), which are armchair leads that extend infinitely in the x-direction. For the zigzag vertical leads $(L_2, L_3, L_5, L_6)$ we can abruptly change the phases so that they now evolve in the x-direction and remain constant in the y-direction, in which the leads are infinite. If the dots and boxes algorithm is followed to match with the previous phase assignations made for the Hall bar, we can make the phases evolve in the x-direction in integer steps of $\theta_H$ with alternating signs, as shown in Figs. 6(a) and 6(e).

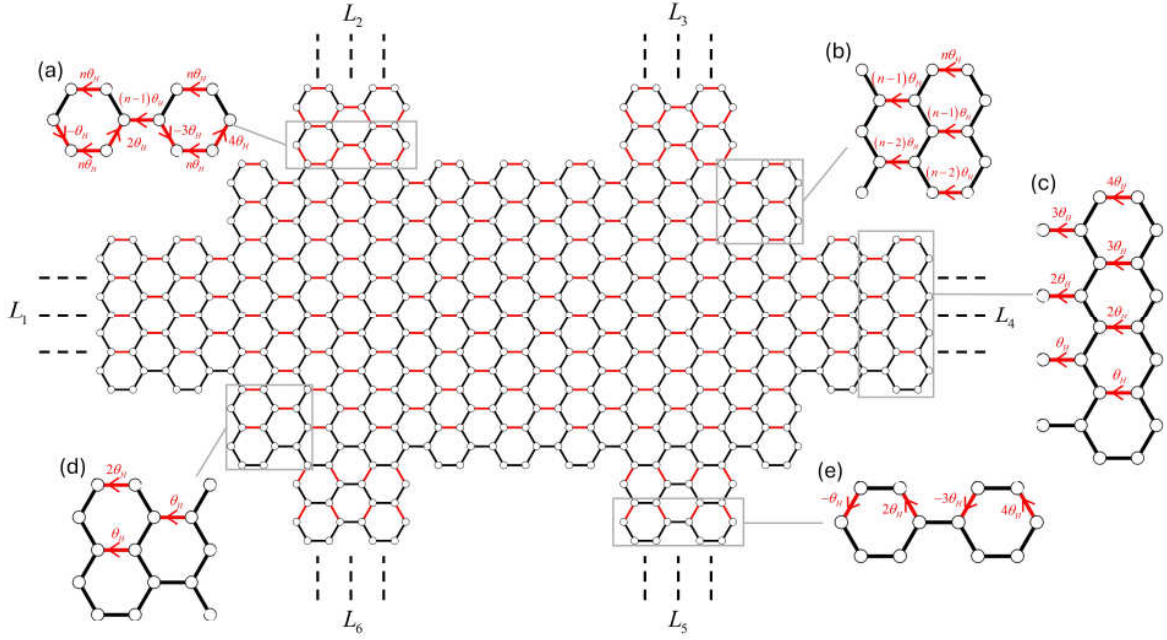

**Figure 6:** Schematic representation of the graphene Hall bar. The black and red lines represent the connections with null and non-zero Peierls phase respectively. We show the Peierls phase assignation for different parts of the Hall bar in (b) and (d) as well as for the leads in (a), (c) and (e).

The Hall bar can be described by a tight binding Hamiltonian of the form

$$\hat{H} = \sum_n \left( \varepsilon_n |n\rangle\langle n| - \sum_{\langle m,n \rangle} t_{n,m} |m\rangle\langle n| \right), \tag{14}$$

where $t_{n,m} = t_0 \exp(i\theta_{n,m})$ with $\theta_{n,m}$ being the Peierls phases we just assigned and $t_0 = -2.78 eV$ is the first neighbors hopping parameter for graphene [37]. The outmost sum develops over all the lattice sites and $\langle m,n \rangle$ denotes that the second sum runs over the $m-$sites that are first neighbors to the $n-$site. The on-site energies $\varepsilon_n$ are set to take random values $\varepsilon_n \in [-W, W]$ to introduce a diagonal disorder into the Hamiltonian, as proposed in Ref. [38]. We refer to $W$ as the disorder strength. It is worth mentioning that we set $\varepsilon_n = 0$ for all the sites in the infinite leads to treat them as periodic systems. In the following we consider a Hall bar that is 200 nm long per 100 nm tall, the leads are 60 nm wide and the separation between the zigzag leads is of 100 nm. The applied magnetic field is of $B = 20\,\text{T}$, which considering a graphene lattice parameter of 0.246 nm [37], results in a flux $\Phi_H = 9.432 \times 10^{-18}$ Wb per unit cell. The calculations were made for a disorder strength of $W = 0.1|t_0|$.

To determine the resistances between the terminals in the Hall bar we can use the Landauer-Büttiker formula [39]

$$I_n = \frac{2e^2}{h} \sum_m \left( T_{m,n}(E) V_n - T_{n,m}(E) V_m \right), \qquad (15)$$

where $I_n$ and $V_n$ are respectively the current and voltage associated to the $n-$th terminal. The transmittance $T_{n,m}(E)$ from the $n-$th terminal into the $m-$th terminal is calculated as a function of the energy $E$ through the RSMM [35,36].

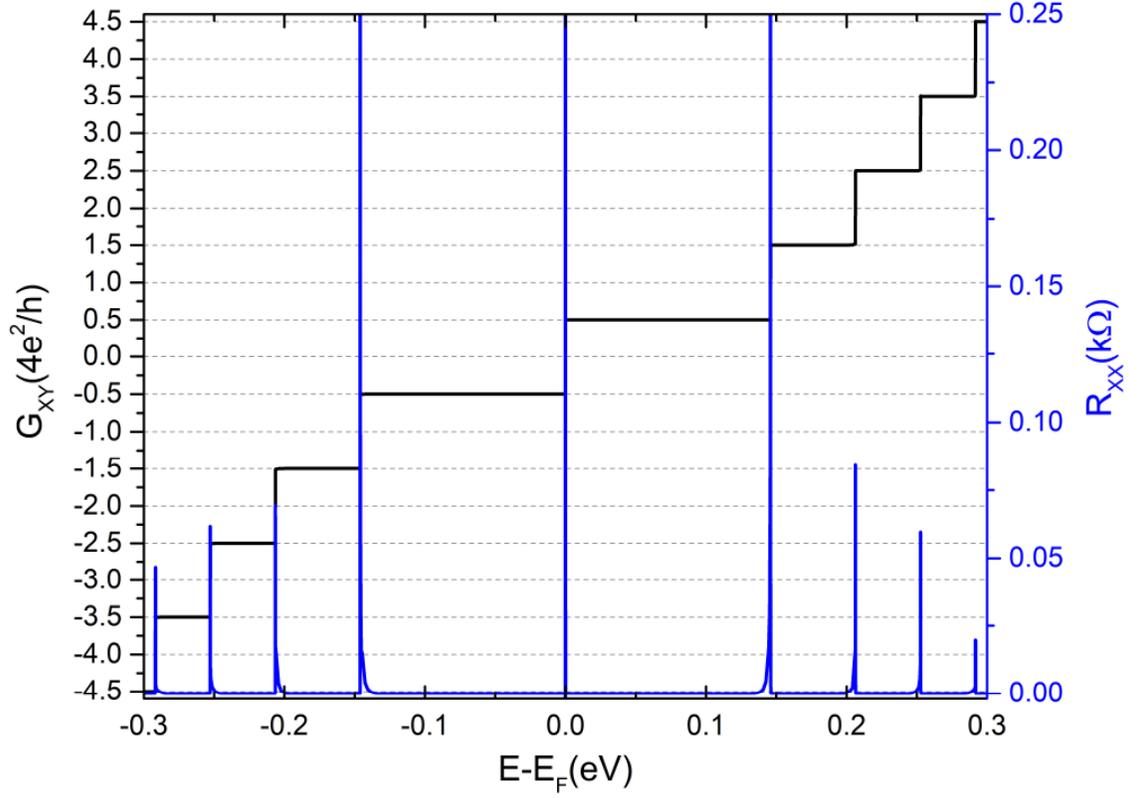

**Figure 7:** Transverse conductance (black) and longitudinal resistance (blue) for a graphene Hall bar modeled by the dots and boxes algorithm.

Motivated by the quantum Hall experiment [1], we can think of a small bias voltage $V_{S,D}$ between the terminals 1 (source) and 4 (drain) and consider the rest of the terminals as floating, that is, such that the net current through them vanish $\left(I_n = 0 \text{ for } n = 2,3,5,6\right)$. Under these conditions, we can substitute into Eq. (15) and solve the equation system to determine the unknown currents and voltages. The transverse Hall conductance $G_{XY}$ can then be obtained from Ohm's Law as

$$G_{XY} = \frac{I_1}{V_6 - V_2}, \tag{16}$$

where $I_1$ is the current flowing between terminals 1 and 4. Likewise, the longitudinal Hall resistance can be calculated as

$$R_{XX} = \frac{1}{G_{XX}} = \frac{V_2 - V_3}{I_1}. \tag{17}$$

Both quantities are shown in Fig. 7. Notice that the transverse Hall conductance is quantized as it takes values of the form $G_{XY} = \frac{2e^2}{h}(2\gamma+1)$ where $\gamma$ is a non-negative integer, which is the expected behavior for the HIQHE [10]. Moreover, we get that the longitudinal resistance $R_{XX}(E)$ is zero except for the Landau levels, where Shubnikov-deHaas oscillations appear due to the high density of bulk states [40].

## 4. Quantum Transport through Domain Walls.

Multidomain Chern insulators have been recently investigated in experimental setups to explore the effects of DWs in electron transport properties [6,18,19]. These experiments have proposed quantum Hall setups as the one shown in Fig. 8 in which the Hall bar in the center has multiple topological domains. The multidomain system can be formed by a non-homogeneous magnetic field that produces different local Chern numbers $(C)$ within the sample. If we think of the terminals as infinite leads it would be realistic to think of those leads as being subjected to the same magnetic field as the regions to which they are attached. This is a relevant aspect as in a Chern insulator as $C$ determines the number of chiral edge states (CES) that appear in the borders of the system. A real-space analysis of such system can be made by a tight-binding model which will allow to describe the system with infinite leads by using methods such as the RSMM or the recursive Green function method. However, the introduction of the non-homogeneous magnetic field results problematic if we were to calculate it directly from Eq. (2) as we would need a vector potential whose curl results in the correct magnetic field for each part of the system and such that the PPs are convenient for the leads oriented in multiple directions. In the following we show how this issue is alternatively solved by the dots and boxes algorithm.

Let us consider a sample formed by a two-dimensional electron gas (2DEG), where the continuous Schrödinger equation can be discretized into a square lattice tight-binding Hamiltonian [41]. This 2DEG description is suitable, for example, for the 2DEG in the AlGa/GaAsAl interphase in which the QHE is reproduced for metrological purposes [42]. Due to the discretization, the hopping parameter amplitude is given as $t_0 = \left(2\hbar^{-2} m_0 a^2\right)^{-1}$, where $m = 0.069 m_0$ is the electron effective mass in the AlGa/AlGaAs interphase and $a$ is the lattice parameter [43]. All the on-site energies take the value $\varepsilon = 4|t_0|$ [41]. If we consider a 2.4 $\mu m$ per 0.8 $\mu m$ sample discretized into a $300 \times 100$ square lattice, the resulting lattice and hopping parameters are $a = 8$ nm and $t_0 = 0.009$ $eV$ respectively.

To achieve multiple topological phases within the sample we divide it in two separate regions, I and II, as shown respectively in red and blue in Fig. 8. We consider magnetic fields in each region such that the fluxes per unit cell are $\Phi_I$ and $\Phi_{II}$ respectively. The system is provided with 6 leads, which are subjected to the same magnetic field as the region in which they are attached. Following the dots and boxes algorithm, we directly assign the PPs in the following manner. For the central rectangular sample as well as for the leads 1 and 4, we can set to zero all the phases associated with vertical bonds. The phases corresponding to horizontal bonds are taken to evolve in multiple steps of $\theta_I = 2\pi \frac{\Phi_I}{\Phi_0}$ and $\theta_{II} = 2\pi \frac{\Phi_{II}}{\Phi_0}$ for the regions I and II, respectively. For the vertical leads (2, 3, 5 and 6) we modify the phase assignation so that the phases are the same for any given unit cell of the lead. This is achieved by keeping the PPs from the horizontal bonds constant while evolving the ones associated to vertical bonds in integer multiples of $\theta_I$ or $\theta_{II}$. This implementation of the algorithm is shown in detail in Fig. 8. Notice how this choice allows to preserve the periodicity of the leads and enables the use of Bloch's theorem.

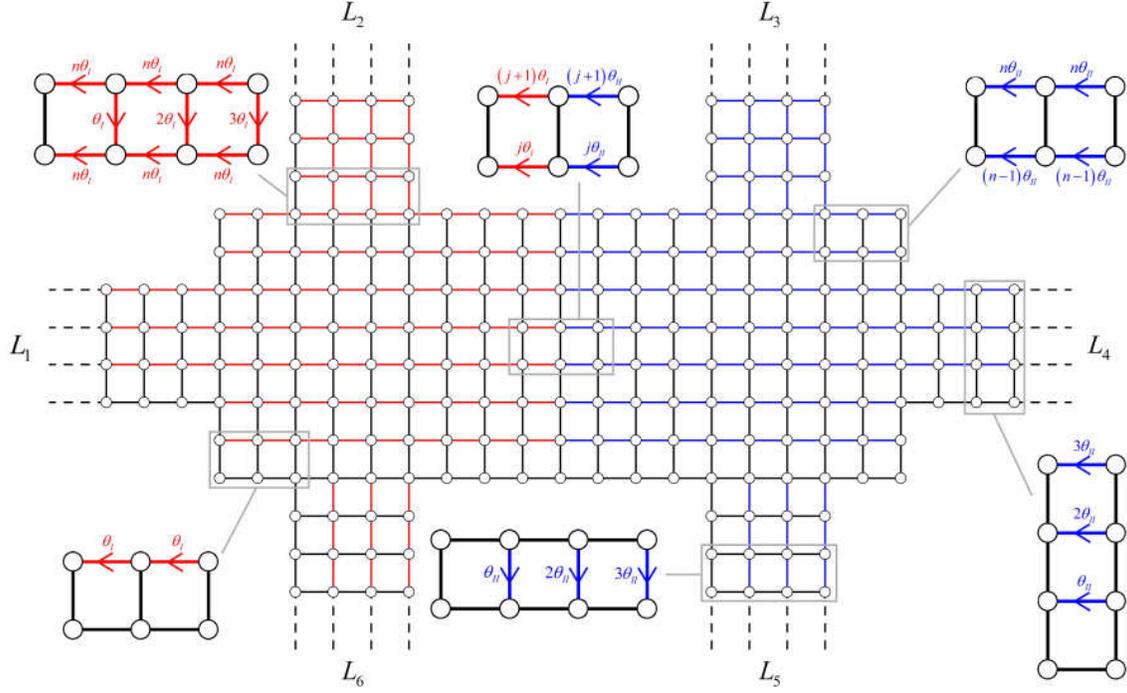

**Figure 8:** Phase assignation for a square lattice with six infinite magnetic leads. The red and blue lines represent non-zero Peierls phases given as integer multiples of the fluxes $\Phi_I$ and $\Phi_{II}$ respectively.

We calculate the resistances between the different terminals by establishing a voltage $V_{S,D}$ between terminals 1 and 4 while setting the rest of the terminals to be floating, as made with the graphene Hall bar in the previous section. The transmittance coefficients $T_{n,m}(E)$ between the terminals are once again calculated by the RSMM, and substituted into the Landauer-Büttiker formula (15) to determine all the voltages and currents in the system. Then, the resistance $R_{n,m}$ between the $n-$th and $m-$th terminals is calculated as

$$R_{n,m} = \frac{V_m - V_n}{I_1}, \qquad (18)$$

where $I_1$ is the current coming out of terminal 1 and running into terminal 4. Let us consider an energy of $E = |t_0|$ and flux $\Phi_I = 0.5\,\hbar/e$, for which the region I has Chern number $C_1 = 1$. By varying the flux $\Phi_{II}$ between $-\Phi_I$ and $\Phi_I$, we calculate the resistance between different terminals as the Chern number evolves in region II. These calculations are shown in Fig. 9. Notice that when $\Phi_I$ has the same sign as $\Phi_{II}$, the resistances are quantized and depend exclusively on the Chern number $C_2$ of region II. However, when the fluxes have different signs, only the transverse resistances $R_{6-2}$ and $R_{5-3}$ are quantized while the longitudinal resistances develop a series of oscillations as a function of energy. This behavior, which to the best of our knowledge has not been previously reported, can be understood as an effect of interference. To confirm this, in the following we calculate the resistance for a non-zero temperature and introduce decoherence into the system.

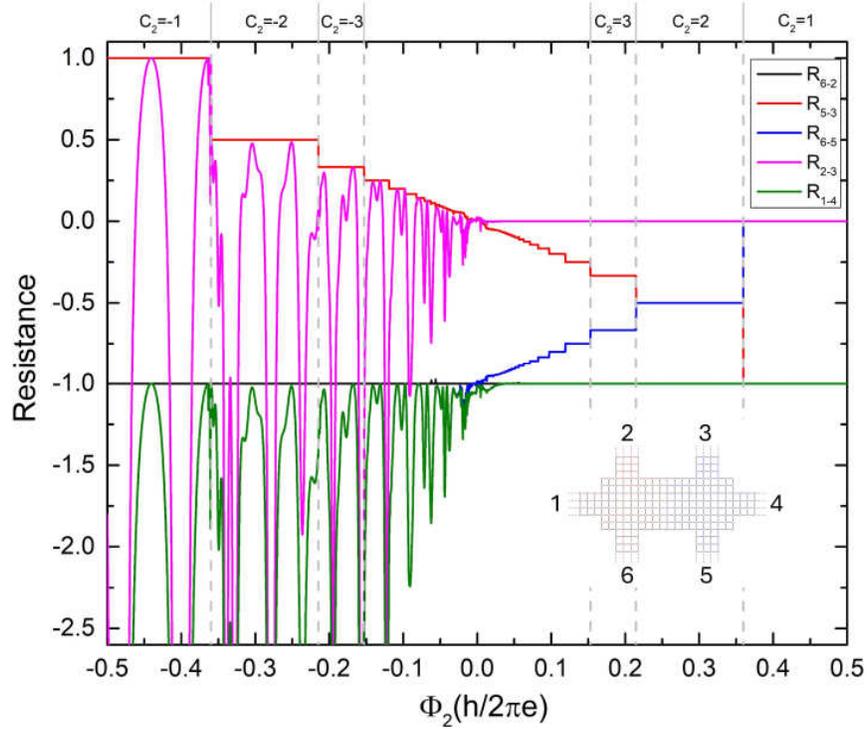

**Figure 9:** Resistance in a quantum Hall multi-terminal setup where two different magnetic fields are applied (as proposed in Fig. 8). The resistance is presented in units of $R_0 = h/2e^2$ and the terminal distribution is shown in the inset.

To include decoherence into the resistance calculations of the Hall bar a set of additional floating leads, known as Büttiker probes, are added to the system. Büttiker probes are a customary tool for adding decoherence into transport calculations [44–46]. The probes are added in terms of a density $\rho \in [0,1]$, that determines the probability that a randomly selected site in the DW has an attached probe. It is worth mentioning that the sites that are connected to Büttiker probes are selected along the DW because far from that region decoherence is not expected to have influence on the border CES, which decay exponentially into the bulk.

On the other hand, the calculation can be made for a finite temperature by using the temperature dependent Landauer-Büttiker formula

$$I_n = \sum_m \left( G_{m,n}(E,T) V_n - G_{n,m}(E,T) V_m \right), \qquad (19)$$

with the conductance

$$G_{n,m}(E,T) = \frac{2e^2}{h} \int_{-\infty}^{\infty} T_{n,m}(E) \left( -\frac{\partial f(E,T)}{\partial E} \right) dE \qquad (20)$$

given in terms of the transmittance and the Fermi distribution $f(E,T) = \left( \exp\left(\frac{E-E_F}{k_B T+1}\right) + 1 \right)^{-1}$, where $k_B$ is the Boltzmann's constant, $E_F$ is the Fermi energy and $T$ is the temperature [39]. For each Büttiker probe connected to the system a new expression given as Eq. (19) is added to the equation system that must be solved to calculate the resistance in the multi-terminal setup. Just as with the rest of the floating terminals, the net current in each Büttiker probe is set to be zero so that there is no net charge flux and their only effect is a phase randomization [46].

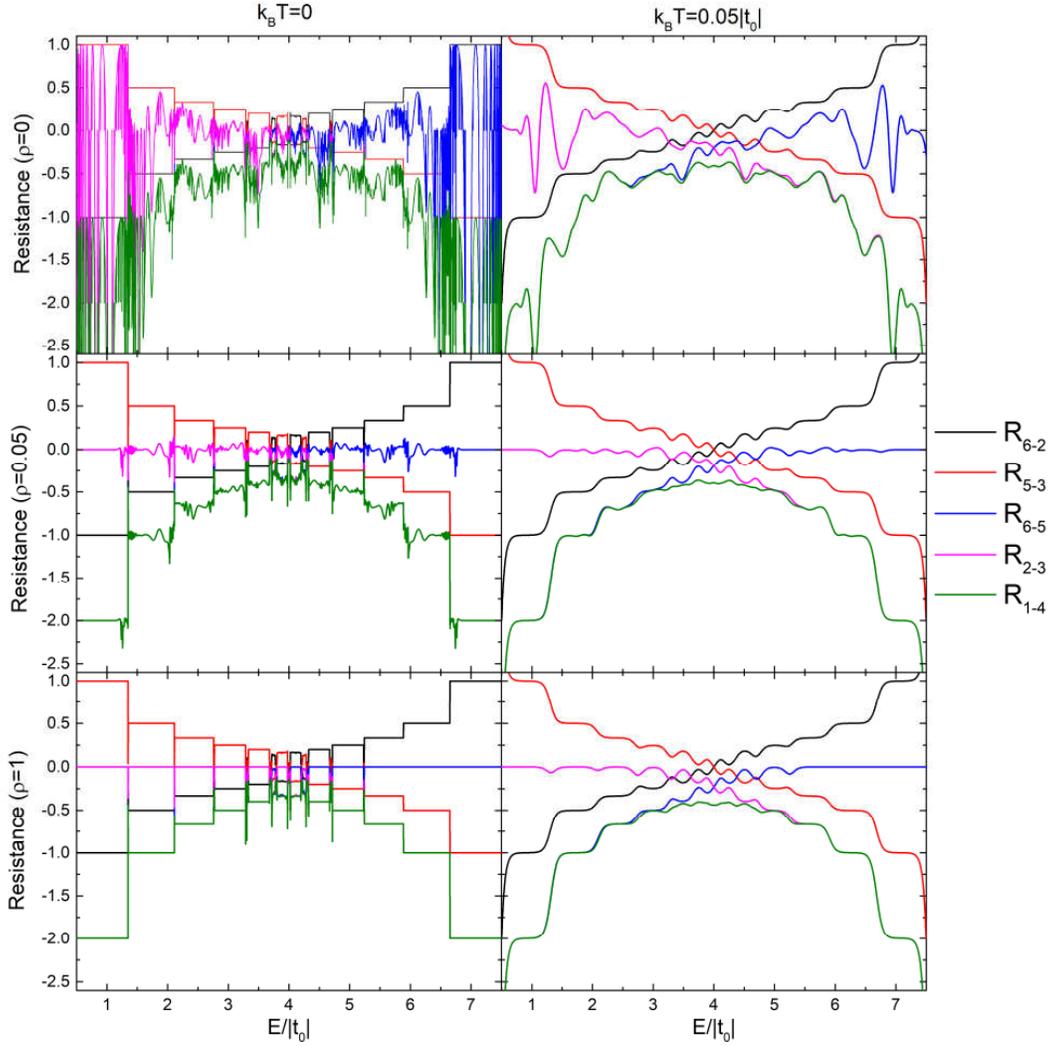

**Figure 10:** Resistance between terminals in the Hall bar with two topological domains when decoherence is added through Büttiker probes. The low row shows calculations made for zero temperature and the right side are calculated for a finite temperature.

The resistances in presence of decoherence and finite temperature are shown in Fig. 10 as a function of the energy. These calculations were made considering magnetic fluxes of $\Phi_I = -\Phi_{II} = 0.5 \frac{\hbar}{e}$. From the graph shown in Fig. 9 we can see that the fluxes are such that the regions $I$ and $II$ have local Chern numbers $C_1 = 1$ and $C_2 = -1$ respectively, and is therefore a case in which the longitudinal resistances oscillate. It can be readily seen in the right side of Fig. 10 that even at zero temperature the presence of decoherence drastically reduces the oscillations, which suggests that they appear due to a coherent phenomenon such as interference. The right column of Fig. 10 shows the decoherent resistances for a thermal energy of $k_B T = 0.05 |t_0|$, which for the abovementioned sample of AsGa/AsGaAl would correspond to a temperature of $T = 5\ K$. Observe how the non-zero temperature softens the oscillations that are already reduced by the decoherence. Under these conditions, the obtained values for each of the resistances are in great agreement with the ones measured in experimental setups [6]. An important remark is that this agreement between the calculations and the experiment occurs only when decoherence and non-zero temperature are both included in the calculations, as neither of these conditions produces a good quantization alone.

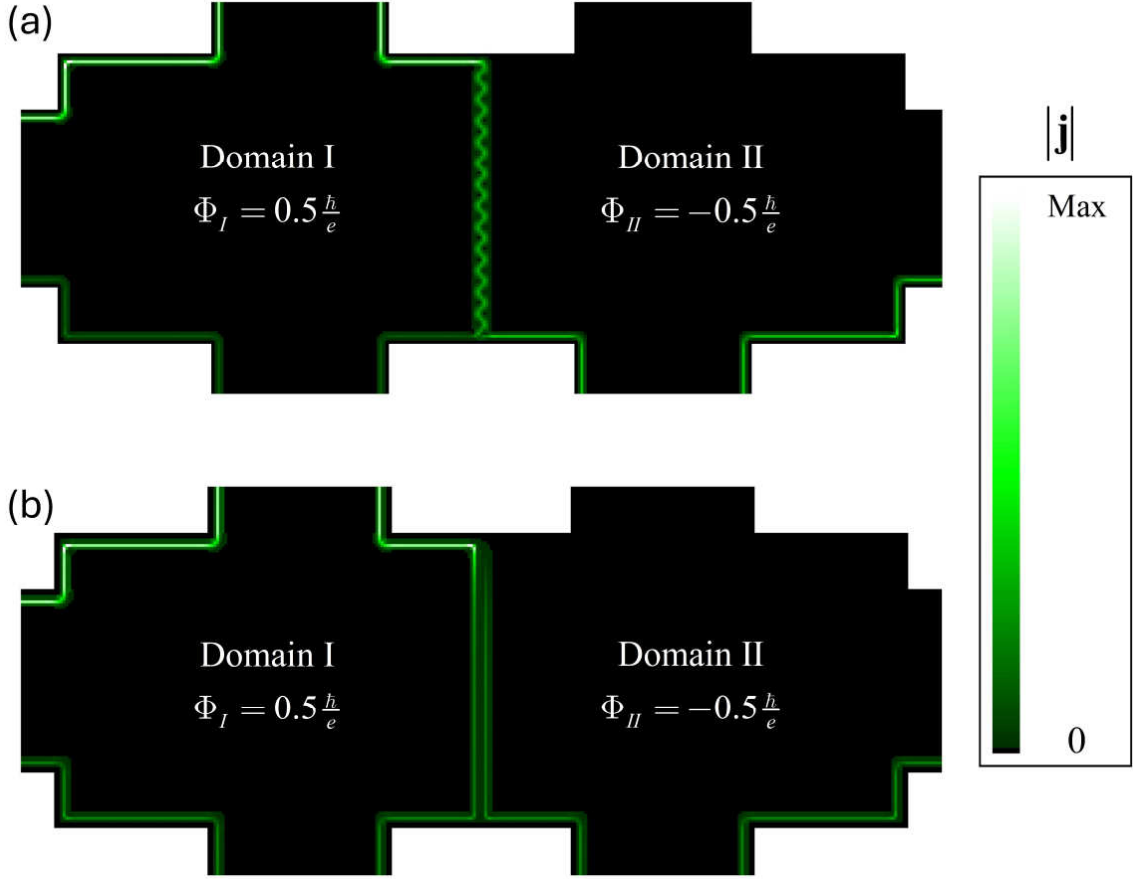

**Figure 11:** Current density calculated for the multidomain Hall bar with magnetic fluxes of $\Phi_I = -\Phi_{II} = 0.5\frac{\hbar}{e}$ for an energy of $E = |t_0|$. (a) Shows the currents for the coherent case and in (b) decoherence is added through Büttiker probes.

An advantage of having a real-space model is that we have access to the wave function and current density around the DW. The current density arises from incoming waves that propagate through open channels in the leads. We calculate the current density on the $n$-th site produced by a wave incoming through the $M$-th open channel ($\mathbf{j}_n^M$) by following the method in Ref. [47]. This approach involves first computing the wavefunctions, which can be done by using the RSMM as described in Ref. [48]. As the total current density is a superposition of the currents calculated for each open channel, the information about the bias voltage to which the Hall bar is subjected can be introduced by weighting the mode specific currents $\mathbf{j}_n^M$ by the voltage of the terminal from which the $M$-th mode arrives into the system. Thus, the total current density is given by

$$\mathbf{j}_n = \sum_M V_M \mathbf{j}_n^M, \tag{21}$$

where the voltages $V_M$ are calculated from the Landauer-Büttiker formula (15) for all the terminals in the Hall bar.

In Fig. 11a we show the current density for the multidomain Hall bar with a magnetic flux of $\Phi_I = -\Phi_{II} = \frac{\hbar}{e}$ and an energy of $E = |t_0|$. For these values there is a single CES in region I(II) that flows clockwise(counter-clockwise). Notice how the current density exhibits a wave-like pattern in the DW, which we refer to as the *current wiggle*. This wiggle is a product of interference that makes the wave-function oscillate. The amount portion of the current that gets transmitted into terminals 5 and 6 depends on how the wiggle arrives at the lower border. This can be controlled by the magnetic field and the energy, resulting in the oscillations shown in Figs. 9 and 10. However, notice from Fig. 11(b) that when decoherence is introduced into the system through Büttiker probes the wiggle disappears and the current density is evenly distributed between terminals 5 and 6, resulting in the resistance values that were previously reported in experiments [6].

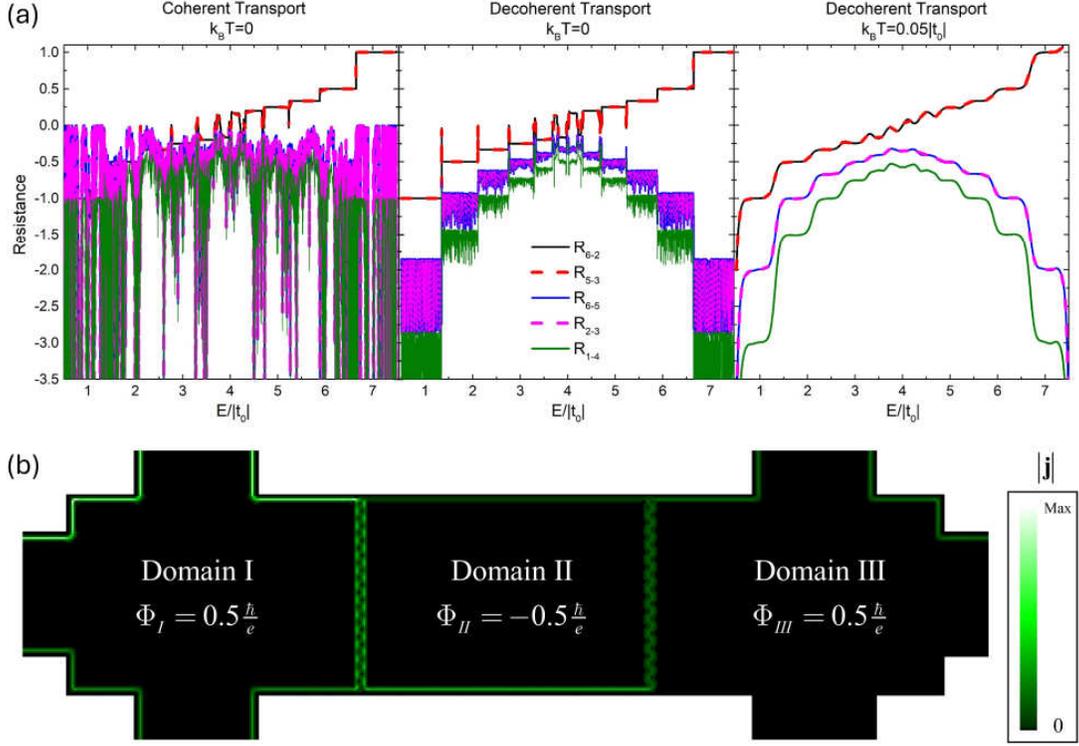

**Figure 12:** (a) Resistance calculated for a Hall bar with two domain walls when the transport is coherent (left panel), with decoherence (middle panel) and for non-zero temperature (right panel). The decoherence is added through Büttiker probes with a density of $\rho = 1$ along the domain walls. (b) Coherent current density in the Hall bar for an energy $E = |t_0|$.

Following the reasoning behind de PP assignation made for the Hall bar shown in Fig. 8, we can extend the analysis to different setups of Hall bars with DWs. For instance, a Hall bar with two DWs can be formed by considering two regions, I and III, that are subjected to the same magnetic flux and are separated by region II which has a magnetic flux with opposite sign. We calculate the resistances in this setup for magnetic fluxes $\Phi_I = -\Phi_{II} = \Phi_{III} = 0.5\frac{\hbar}{e}$. The calculations are made for a Hall bar modeled as a square lattice with $N = 450$ sites long per $M = 100$ sites tall. Each domain has 150 sites long per 100 sites tall. The resulting resistances are shown in Fig. 12(a). Just as in the Hall bar with a single DW, the resistances for the coherent case exhibit important oscillations as the energy is varied (left panel). These oscillations are mitigated when Büttiker probes are added in the DWs (medium panel) and are averaged into a well-defined quantization for non-zero temperature (right panel). The quantized decoherent resistances obtained for non-zero temperature are also in great agreement with the experimental values reported in Ref. [6]. The current density corresponding to the coherent case is presented in Fig. 12(b). Obseve that the current form interference wiggles in both DWs, which shows that the oscillations for the coherent calculation of the resistances have the same nature as the ones discussed in the single DW case.

## 4  Conclusions

In this paper we have shown that the calculation of Peierls phases (PP) can be done without an explicit calculation of the vector potential in discrete systems. In fact, the PPs can be assigned in any convenient way given that Eq. (10), which is the only relevant constraint established by the gauge-invariant magnetic flux, is satisfied. These ideas are synthesized in a graphic method that we call the dots and boxes algorithm that allows to identify the PPs that can be freely assigned and the ones that must take specific values to satisfy Eq. (10). To show how the method can be implemented and as a validation of it, we calculated the transmittance of an Aharonov-Bohm ring resulting in the characteristic periodic behavior as a function of the magnetic flux with period $\Phi_0$. This analysis was taken further by adding a central chain that divides the flux in two parts $\Phi_I$ and $\Phi_{II}$, and shown that the transmittance is still periodic. The algorithm was also applied to make a non-conventional PP assignation in a graphene quantum Hall bar, where the resistance between terminals was calculated and resulted in the values expected for the Half-Integer Quantum Hall Effect.

In the last section, the dots and boxes algorithm is utilized to describe a Hall bar formed by a multi-domain Chern insulator. This was achieved by assuming a non-homogeneous magnetic field that induces two different local Chern numbers within the bar. It was found that whenever the local Chern numbers have opposite signs, the longitudinal resistances were not quantized but exhibited big oscillations. To show that these oscillations appear due to interference, decoherence was added to the system by Büttiker probes in the domain wall (DW). Once decoherence was included it was found that at finite temperature the oscillations disappear and all the resistances in the Hall bar show a great agreement with the quantization reported in previous experiments [6].

**Acknowledgements**

This work was supported by UNAM-PAPIIT IN109022. Ricardo Y. Diaz-Bonifaz thanks CONAHCYT for the graduate scholarship granted. Computations were performed at Miztli under project LANCAD-UNAM-DGTIC-329.